\documentclass[twocolumn]{aastex631}

\usepackage[utf8]{inputenc}
\DeclareUnicodeCharacter{2212}{\textminus}
\shorttitle{The Abundance of GD-1}
\shortauthors{Zhao et al.}
\graphicspath{{./}{figures/}}

\begin{document}

\title{Chemical abundances of seven stars in {\bf the} GD-1 stream }

\correspondingauthor{Jingkun Zhao; Guangwei Li; Wako Aoki; Gang Zhao}
\email{zjk@bao.ac.cn; lgw@nao.cas.cn; aoki.wako@nao.ac.jp; gzhao@nao.cas.cn}

\author[0000-0003-2868-8276]{Jing-Kun Zhao}
\affiliation{ National Astronomical Observatories, Chinese Academy of Sciences, Beijing 100101, People's Republic of China},
\author{Guang-Wei Li}
\affiliation{ National Astronomical Observatories, Chinese Academy of Sciences, Beijing 100101, People's Republic of China},
\author{Wako Aoki}
\affiliation{National Astronomical Observatory of Japan
2-21-1 Osawa, Mitaka, Tokyo 181-8588, Japan}
\affiliation{Astronomical Science Program, The Graduate University for Advanced Studies, SOKENDAI, 2-21-1 Osawa, Mitaka, Tokyo 181-8588, Japan},
\author[0000-0002-8980-945X]{Gang Zhao}
\affiliation{ National Astronomical Observatories, Chinese Academy of Sciences, Beijing 100101, People's Republic of China}
\affiliation{School of Astronomy and Space Science, University of Chinese Academy of Sciences, Beijing 100049, People's Republic of China},
\author{Guo-Chao Yang}
\affiliation{School of Physics and Astronomy, China West Normal University, 637009 Nanchong, People's Republic of China},
\author{Jian-Rong Shi}
\affiliation{ National Astronomical Observatories, Chinese Academy of Sciences, Beijing 100101, People's Republic of China},
\author{Hai-Ning Li}
\affiliation{ National Astronomical Observatories, Chinese Academy of Sciences, Beijing 100101, People's Republic of China},
\author{Tadafumi Matsuno}
\affiliation{Astronomisches Rechen-Institut, Zentrum für Astronomie der Universit\"{a}t Heidelberg, M\"{o}nchhofstraße 12-14, 69120 Heidelberg, Germany},
\author{Miho Ishigaki}
\affiliation{National Astronomical Observatory of Japan
2-21-1 Osawa, Mitaka, Tokyo 181-8588, Japan},
\author{Takuma Suda}
\affiliation{Department of Liberal Arts, Tokyo University of Technology, Nishi Kamata 5-23-22, Ota-ku, Tokyo 144-8535, Japan}
\affiliation{Research Center for the Early Universe, The University of Tokyo, 7-3-1 Hongo, Bunkyo-ku, Tokyo 113-0033, Japan},
\author{Satoshi Honda}
\affiliation{Nishi-Harima Astronomical Observatory, Center for Astronomy, University of Hyogo, 407-2, Nishigaichi, Sayo-cho, Sayo, Hyogo 679-5313, Japan},
\author{Yu-Qin Chen}
\affiliation{ National Astronomical Observatories, Chinese Academy of Sciences, Beijing 100101, People's Republic of China},
\author{Qian-Fan Xing}
\affiliation{ National Astronomical Observatories, Chinese Academy of Sciences, Beijing 100101, People's Republic of China},
\author[0000-0002-8609-3599]{Hong-Liang Yan}
\affiliation{ National Astronomical Observatories, Chinese Academy of Sciences, Beijing 100101, People's Republic of China},
\author[0000-0001-7609-1947]{Yong Yang}
\affiliation{ National Astronomical Observatories, Chinese Academy of Sciences, Beijing 100101, People's Republic of China}
\affiliation{School of Astronomy and Space Science, University of Chinese Academy of Sciences, Beijing 100049, People's Republic of China},
\author{Xian-Hao Ye}
\affiliation{ National Astronomical Observatories, Chinese Academy of Sciences, Beijing 100101, People's Republic of China},



\begin{abstract}
We present the first detailed chemical abundances for seven GD-1 stream stars from Subaru/HDS spectroscopy. Atmospheric parameters were derived via color calibrations ($T\rm_{eff}$) and iterative spectroscopic analysis. LTE abundances for 14 elements ($\alpha$, odd-Z, iron-peak, n-capture) were measured. Six stars trace the main orbit, one resides in a `blob'. All exhibit tightly clustered metallicities ([Fe/H] = -2.38, {\bf intrinsic dispersion smaller than 0.05 dex, average uncertainty is about 0.13 dex}). While one star shows binary mass transfer signatures, the other six display consistent abundance patterns (dispersions $<$ uncertainties). Their iron-peak elements (Sc, Cr, Mn, Ni) match Milky Way halo stars. In contrast, Y and Sr are systematically lower than halo stars of similar [Fe/H]. Significantly, six stars show consistently enhanced [Eu/Fe] $\sim$ 0.60 ($\sigma$ = 0.08). A tight Ba-Eu correlation (r = 0.83, p=0.04) exists, with [Ba/Fe] = -0.03 $\pm$ 0.05, indicating a common r-process origin. This extreme chemical homogeneity strongly supports an origin from a single disrupted globular cluster. The lack of light-element anti-correlations may stem from our sample size or the progenitor's low mass.

\end{abstract}

\keywords{Milky Way Galaxy (1054) --- Chemical abundances(224) --- Tidal tails (1701) --- Globular star clusters (656)}


\section{Introduction} \label{sec:intro}
The GD-1 stellar stream was first detected using photometric data from the Sloan Digital Sky Survey (SDSS; \citealt{2000AJ....120.1579Y}) Data Release 4 \citep{2006ApJ...643L..17G}. Subsequent studies significantly improved both its sky coverage and photometric uniformity \citep{2009ApJ...697..207W,2010ApJ...712..260K}. Currently, over 811 member stars with high probability of association with the GD-1 stellar stream  have been identified \citep{2021ApJ...914..123I}.

This remarkable stellar stream extends over $80^\circ$ across the northern sky, passing within $30^\circ$ of the Galactic pole. The visible portion lies at a median distance of $\sim$8 kpc, with distances ranging from approximately 7 to 11 kpc along its length. With an angular width of merely {\bf $0.4^\circ$} (corresponding to a linear width of about 70 pc), GD-1 displays an extraordinary length-to-width ratio exceeding 100:1.

Collisional N-body simulations by \cite{2019MNRAS.485.5929W} suggest that GD-1's progenitor was likely a very low-mass GC (a few 10$^{4}$ M$_\odot$) that dissolved within the last 3 Gyr. These findings place GD-1's progenitor at the low-mass extreme of Milky Way GCs, with a dissolution timescale significantly shorter than a Hubble time. This rapid dissolution timescale may indicate either an extragalactic origin or atypical formation and evolutionary processes for this GC.

The GD-1 stellar stream has emerged as a powerful dynamical probe for constraining the Milky Way's gravitational potential, owing to its exceptionally thin and coherent orbit. Early constraints were established by \cite{2010ApJ...712..260K} using a six-dimensional phase-space map of GD-1, yielding measurements of the circular velocity (Vc = 221 $\pm$ 18 km $\rm s^{-1}$) at the solar radius and potential flattening  (q$\phi$ = 0.89 at 90\% confidence) at Galactocentric radii near R $\sim$ 15 kpc. Subsequent analysis by \cite{2016ApJ...833...31B} determined the overall potential flattening of  the inner halo of Milky Way to be 0.95$\pm$0.04 with GD-1 stream. Using ESA/Gaia astrometry \citep{2018A&A...616A...1G,2018A&A...616A..10G} together with SEGUE \citep{2009AJ....137.4377Y}  and LAMOST measurements \citep{2006ChJAA...6..265Z,2012RAA....12..723Z} of the GD-1 stellar stream, \cite{2019MNRAS.486.2995M} found that the orbital solutions for GD-1 require the circular velocity at the solar radius V$\rm_{circ}$ to be around 244 $\pm$ 4 km s$\rm^{-1}$, and also that the density flattening of the dark halo q about 0.82$^{+0.25}_{-0.13}$. The corresponding Galactic mass within 20 kpc was estimated to be M$\rm_{MW}$( $<$ 20{ kpc}) = 2.5 $\pm$ 0.2 $\times$ 10$^{11}$  M$_{\odot}$. {\bf \cite{2025ApJ...985L..22N} measured the Galactic acceleration field along the GD-1 stream through phase-space track differentiation, deriving two  Galactic parameters: (1) an enclosed mass of  1.4 $\pm$ 0.1 $\times$ 10$^{11}$  M$_{\odot}$  within 14 kpc, and (2) a z-axis density flattening  q about 0.81$^{+0.06}_{-0.03}$. 

The GD-1 stream serves as a sensitive probe of dark matter substructure, particularly through detection of velocity perturbations. Its potential origin as a disrupted low-mass globular cluster enhances sensitivity to such perturbations. \citet{2019ApJ...880...38B} developed a comprehensive model demonstrating how interactions with massive perturbers can reproduce observed stream features, including gaps and spurs.  Recent work by \cite{2025arXiv250313290C} showed that continuous heating by an evolving CDM subhalo population over $\sim$ 11 Gyr can produce a velocity dispersion of 6.2$\pm$1.7 km s$\rm^{-1}$. \citep{2025ApJ...983...68N} forecasted the velocity dispersion of the GD-1 stream, and found that observations are in agreement with
 a CDM subhalo population, with a slight preference for more dense subhalos. }

The metallicity properties of the GD-1 stellar stream have been extensively investigated through multiple observation data. Using spectroscopic data from the SDSS/SEGUE \citep{2009AJ....137.4377Y}, \citet{2009ApJ...697..207W} gave an average metallicity for the GD-1 stream of [Fe/H] =$ -2.1$ $\pm$ 0.1 and an age comparable to the GC M92. On the other hand, a relatively more metal-rich
([Fe/H] = $-1.4$) and younger age (9 Gyr) for the GD-1 stream
are found by \citet{2010ApJ...712..260K}, based on the isochrone
fitting to the photometric data. \citet{2019ApJ...877...13H} obtain the average metallicity of GD-1 stream [Fe/H] $\sim$ -1.96 using 67 member candidates from SDSS/SEGUE \citep{2009AJ....137.4377Y} and LAMOST spectroscopic data \citep{2006ChJAA...6..265Z,2012RAA....12..723Z}.
\citet{2018ApJ...869..122L} derived the [Fe/H] $\sim$ -2.2 of GD1 stream combining data from Gaia DR2, SDSS DR14, and LAMOST DR6. 
\citet{2020ApJ...892L..37B} show the GD-1 stream has very little spread in the [Fe/H] and [$\alpha$/Fe] distribution and the mean value of [Fe/H] is about -2.3. This chemical uniformity strongly suggested the stream originated from a single, well-mixed progenitor system.

Despite extensive studies of the GD-1 stellar stream, a detailed chemical abundance analysis of multiple elements has remained elusive. In this paper, we present the first high-resolution spectroscopic study of seven GD-1 member stars, reporting their abundance patterns to shed new light on the stream’s progenitor.

The paper is structured as follows: Section 2 describes the observations and data reduction procedures. Section 3 details our methodology for abundance determination. In Section 4, we present and discuss the results, and Section 5 summarizes our key findings and their implications.

\begin{deluxetable*}{cccccccc}
	\tablenum{1}
	\tablecaption{Basic Parameters of the Seven Stars and the Subaru/HDS Observation \label{tab:Objects}}
	\tablewidth{0pt}
	\tablehead{
		\colhead{No} & \colhead{ID} & \colhead{RA} & \colhead{DEC} &
		\colhead{G} & \colhead{Date} & \colhead{exposure time}&\colhead{S/N}$\tablenotemark{a}$ \\
		\colhead{} & \colhead{} & \colhead{(degree)} & \colhead{(degree)} &
		\colhead{(mag)} & \colhead{} &\colhead{(sec)}& \colhead{}
	}
	\decimalcolnumbers
	\startdata
	A & J0932+2841 &143.176219  & 28.684163 & 14.565 & 2020 March 11 & 2400*3 &101 \\
	B & J0930+2902 &142.671178 & 29.032959 & 14.285 &2020 March 11 & 2400*3 & 83 \\
	C &J0955+3553 & 148.804203 & 35.888585 &13.434 & 2020 March 11 &  1200*3 & 67 \\
	D & J1015+4213 & 153.787448 & 42.230355 & 13.270 & 2020 March 11 & 1200*3 & 101\\
	E & J1048+4708 & 162.052801 &47.145379 & 14.466 &2020 March 11 & 2400*3 &79 \\
	F & J1213+5559 & 183.379156 & 55.988948 & 13.359 & 2020 March 11 & 1200*3 & 38\\
	G & J1358+5826 &209.542437 & 58.440025 &13.023 & 2020 March 11 & 1200*3 &43 \\
	\enddata
	\tablenotetext{a}{The S/N per pixel was measured at $\lambda \sim$ 5170 {\AA}. }
\end{deluxetable*}

\begin{deluxetable*}{ccccccccccccccc}
	\tablenum{2}
	\tablecaption{Stellar Parameters of the Six Stars \label{tab:Parameter}}
	\tablewidth{0pt}
	\tablehead{
		\colhead{ID} & \colhead{g} & \colhead{r} & \colhead{i} & \colhead{Ks} & \colhead{ebv} &\colhead{$T\rm_{eff}$}& \colhead{log $g$}&\colhead{[Fe/H]} & \colhead{$\xi_{t}$}& \colhead{Vr} & \colhead{$T\rm_{eff}$\_l} & \colhead{log $g$\_l} &
		\colhead{[Fe/H]\_l}   & \colhead{Vr\_Gaia}\\
		\colhead{} &\colhead{(mag)} &\colhead{(mag)} &\colhead{(mag)} &\colhead{(mag)} &\colhead{} & \colhead{(K)} & \colhead{(dex)} & \colhead{(dex)} &  \colhead{(km $\rm s^{-1}$)} &\colhead{(km $\rm s^{-1}$)}&
		\colhead{(K)} & \colhead{(dex)} &\colhead{(dex)}& \colhead{(km $\rm s^{-1}$)}
	}
	\decimalcolnumbers
	\startdata
      J0932+2841 &15.07 & 14.55&14.31&12.66&0.020& 5010  & 2.33 & -2.30 & 1.80 & 49.80 &  5104&1.827&-2.349&52.96 \\
      J0930+2902 &14.75 & 14.27&14.07&12.57&0.019& 5296  & 2.3 & -2.36 & 1.89 & 46.4 &--  &--&--&--\\ 
      J0955+3553 &13.98 & 13.41&13.19&11.42&0.013& 4844  & 1.50 & -2.42 & 2.20 & -13 & -- &--&--&-12.43 \\
	  J1015+4213 &13.89 & 13.31&13.02&11.13&0.015& 4651  & 1.50 & -2.35 & 1.90 & -62.8 &  4544&1.101&-2.33&-61.71 \\
	  J1048+4708 &14.99 & 14.45&14.20&12.46&0.021& 4882  & 1.94 & -2.35 & 1.50 & -114.4 &  4918&1.937&-2.191&-111.23\\
	  J1213+5559 &13.96 & 13.32&13.08&11.25&0.012& 4726  & 1.59 & -2.42 & 2.10 & -203.4 &  4578&1.194&-2.394&-206.63 \\
	  J1358+5826 &13.67 & 13.04&12.76&10.81&0.008& 4571  & 1.27 & -2.46 & 1.60 & -280.6 &  4637&1.172&-2.232&-279.27 \\
\enddata
\end{deluxetable*}

\section{Observation and data reduction} \label{sec:obervation}
\subsection{Sample selection}
\begin{figure*}
	\includegraphics[width=\linewidth]{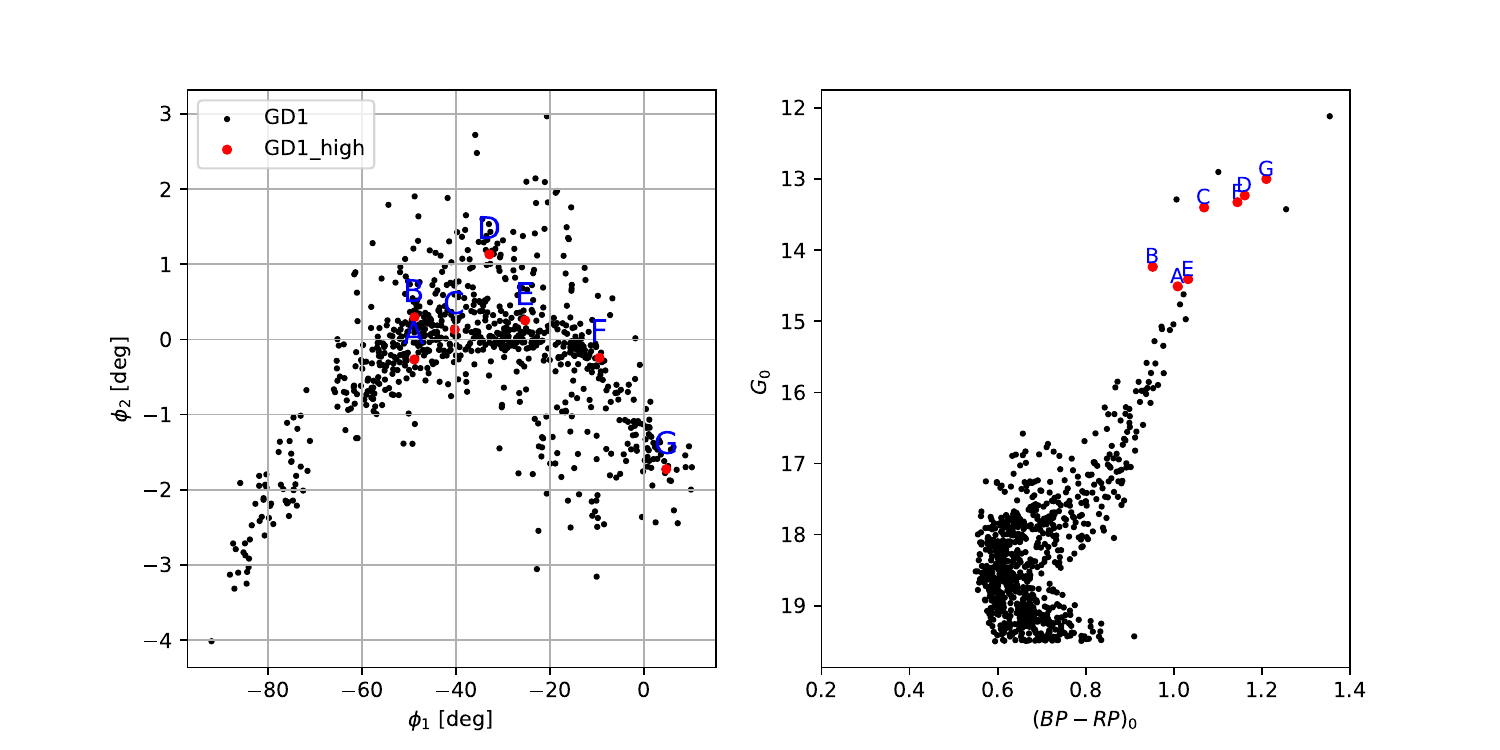}
	\caption{Orbit and CMD of GD-1 stream. Black points are members from \citet{2021ApJ...914..123I} while red points are our seven program stars
		\label{fig:cmd}}
\end{figure*}

Our program stars are selected from candidate member stars of the GD-1 stream using spectra data from LAMOST \citep{2012RAA....12..723Z} and SEGUE \citep{2009AJ....137.4377Y} combined with Gaia DR3 \citep{2023A&A...674A...1G}. The method of identifying members is from \citet{2018ApJ...869..122L}. 
{\bf Among these, we selected some bright stars for high resolution follow-up spectroscopy observation with Subaru/HDS. Due to limited telescope allocation (1 night), we successfully observed the seven brightest stars in this subsample.}
Figure \ref{fig:cmd} shows the orbit and the color magnitude diagram (CMD) of GD-1 stream. The black points represent the member candidates of the GD-1 stream {\bf identified by \citet{2021ApJ...914..123I} using Gaia data alone}. Red points marked with `A' to `G' are seven program stars.  The location of the program stars in $\phi_1$ vs. $\phi_2$ is displayed in the left panel of Figure \ref{fig:cmd}. It should be noticed that star `D' is in blob substructure while others in the main orbits of this stream.

{\bf All observed stars are in late evolutionary stages, as evidenced by their stellar parameters. However, none have reached the asymptotic giant branch (AGB) phase. This evolutionary status confirms that s-process nucleosynthesis has not contributed to their neutron-capture element abundances. Our high-resolution spectroscopic analysis yielded precise measurements for 14 elements spanning four nucleosynthetic groups: (1) $\alpha$-elements (O, Mg, Si, Ca, Ti), (2) odd-Z elements (Na), (3) iron-peak elements (Sc, Cr, Mn, Ni), and (4) n-capture elements (Sr, Y, Ba, Eu). This comprehensive coverage ensures our abundance patterns are free from selection biases.}
 
\subsection{Subaru/High Dispersion Spectrograph Observations}
Observations of the seven target stars were carried out in March 2020 using the High Dispersion Spectrograph (HDS; \citealt{2002PASJ...54..855N}) with the standard StdYd setup, providing a wavelength coverage of 4000-6800 $\rm\AA$ at a resolution power of approximately 36,000. The observational details for these stars, including their identification numbers, coordinates, G magnitudes, observation dates, exposure times, and signal-to-noise ratios (SNR) at $\lambda$ $\sim$ 5170 $\rm\AA$, are presented in Table \ref{tab:Objects}. Data reduction was performed using the IRAF echelle package, incorporating bias-level correction, scattered light subtraction, flat-fielding, spectral extraction, and wavelength calibration with Th-Ar arc lines, with cosmic-ray removal following the method of \citet{2005ApJ...632..611A}. Radial velocities were determined through cross-correlation between the observed spectra and a solar atlas using the HDSV code (\citealt{2007PNAOC...4..153Z}), achieving uncertainties below 1.0 km $\rm s^{-1}$, while continuum normalization was accomplished by fitting low-order polynomials to selected spectral windows.

\begin{figure*}
    \includegraphics[width=\linewidth]{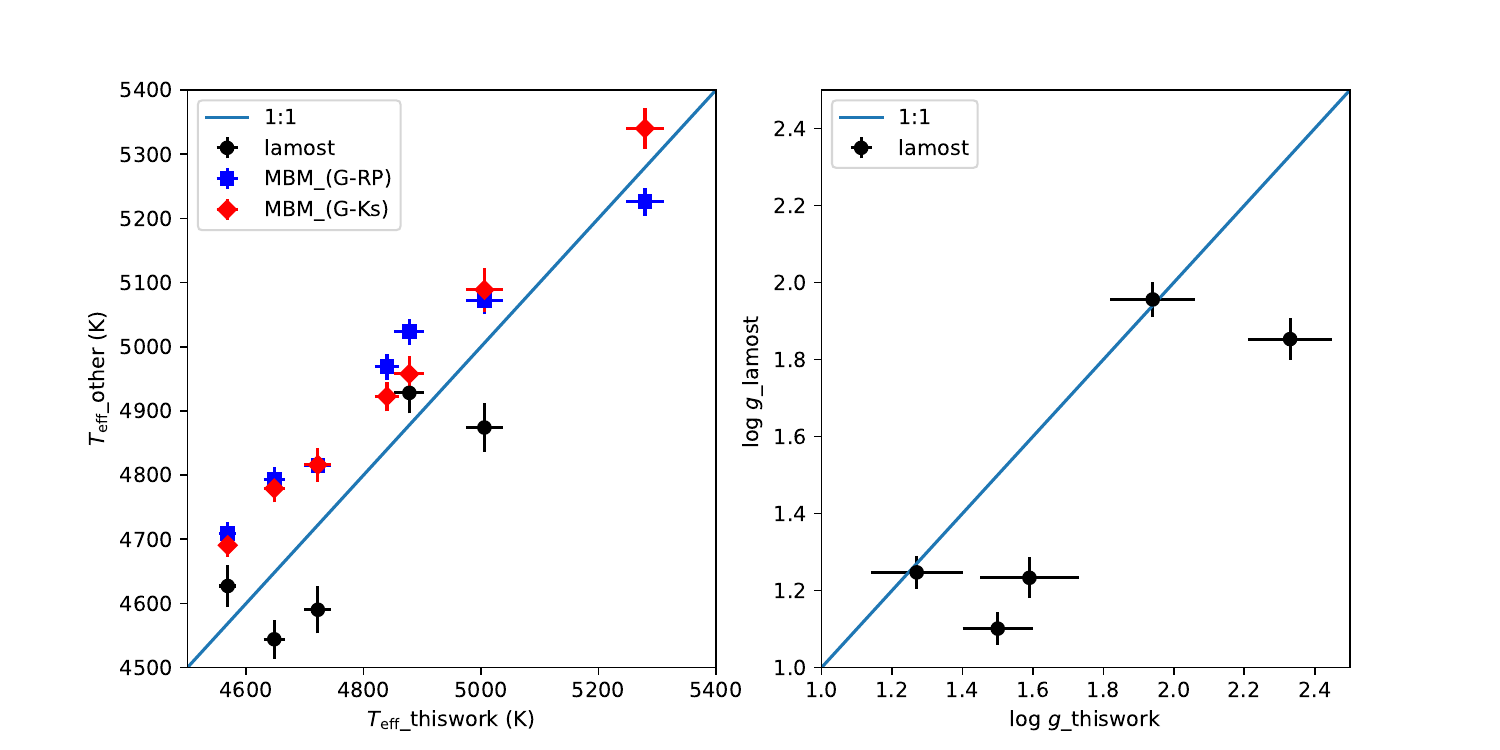}
	\caption{Left: 
    Comparison of stellar parameters between this work and reference values. Left: Effective temperatures ($T\rm_{eff}$) derived in this work versus values from the LAMOST pipeline (black points), MBM (G-G$\rm_{RP}$) calibrations (blue points), and MBM (G-Ks) calibrations (red points). Right: Surface gravity (log $g$) comparison between our spectroscopic determinations and LAMOST pipeline values. Solid lines in both panels indicate the one-to-one relations.
     \label{fig:para_cmp}}
\end{figure*}

\begin{figure*}
	    \includegraphics[width=\linewidth]{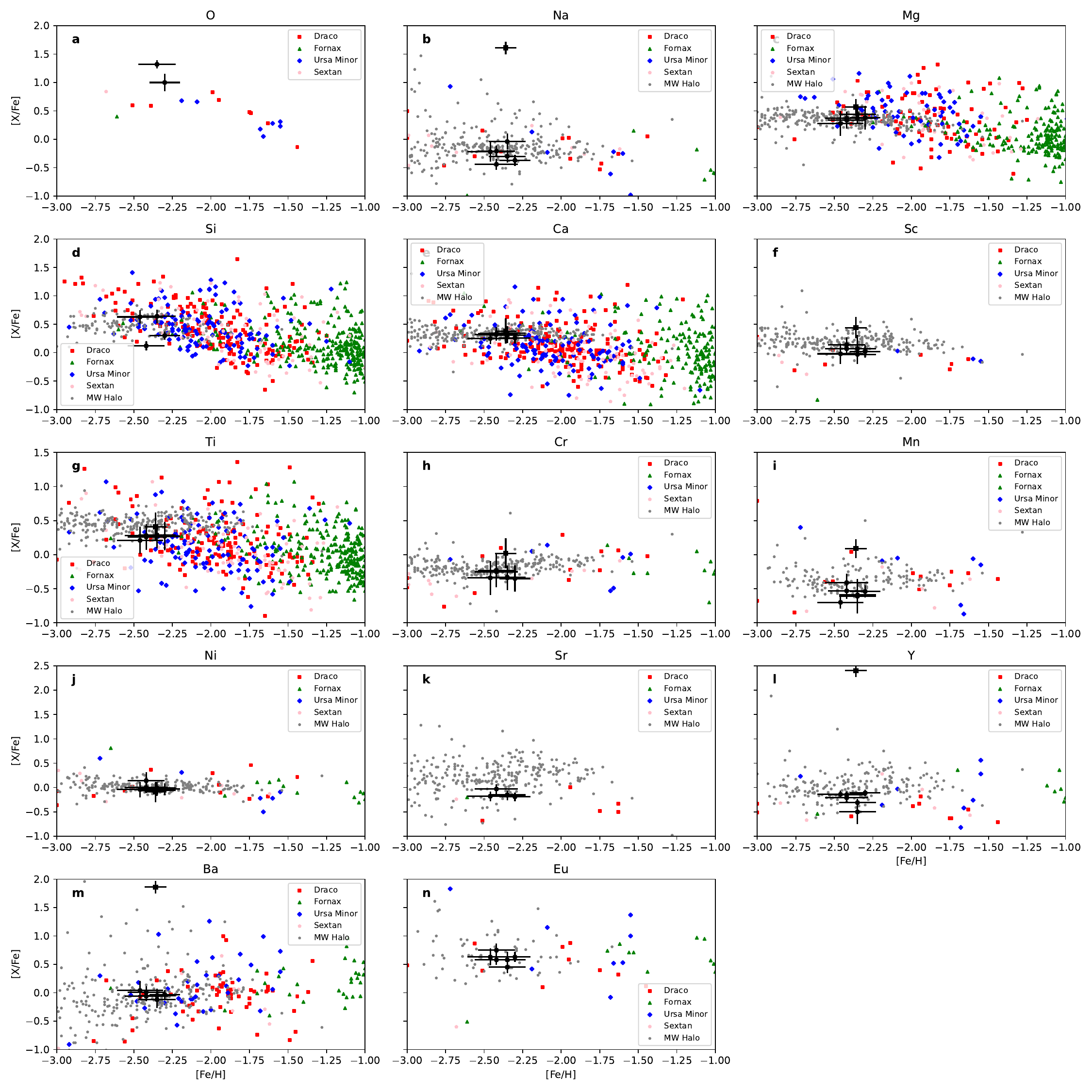}
	\caption{Abundance pattern of seven member stars (black points for six stars and squares for carbon star )  and comparison with dwarf galaxies and the halo stars of Milky Way.  Gray points are metal-poor stars from \cite{2022ApJ...931..147L}. Colored symbols are from dwarf galaxies.
		\label{fig:abu-dis}}
\end{figure*}

\begin{figure*}
     \includegraphics[width=\linewidth]{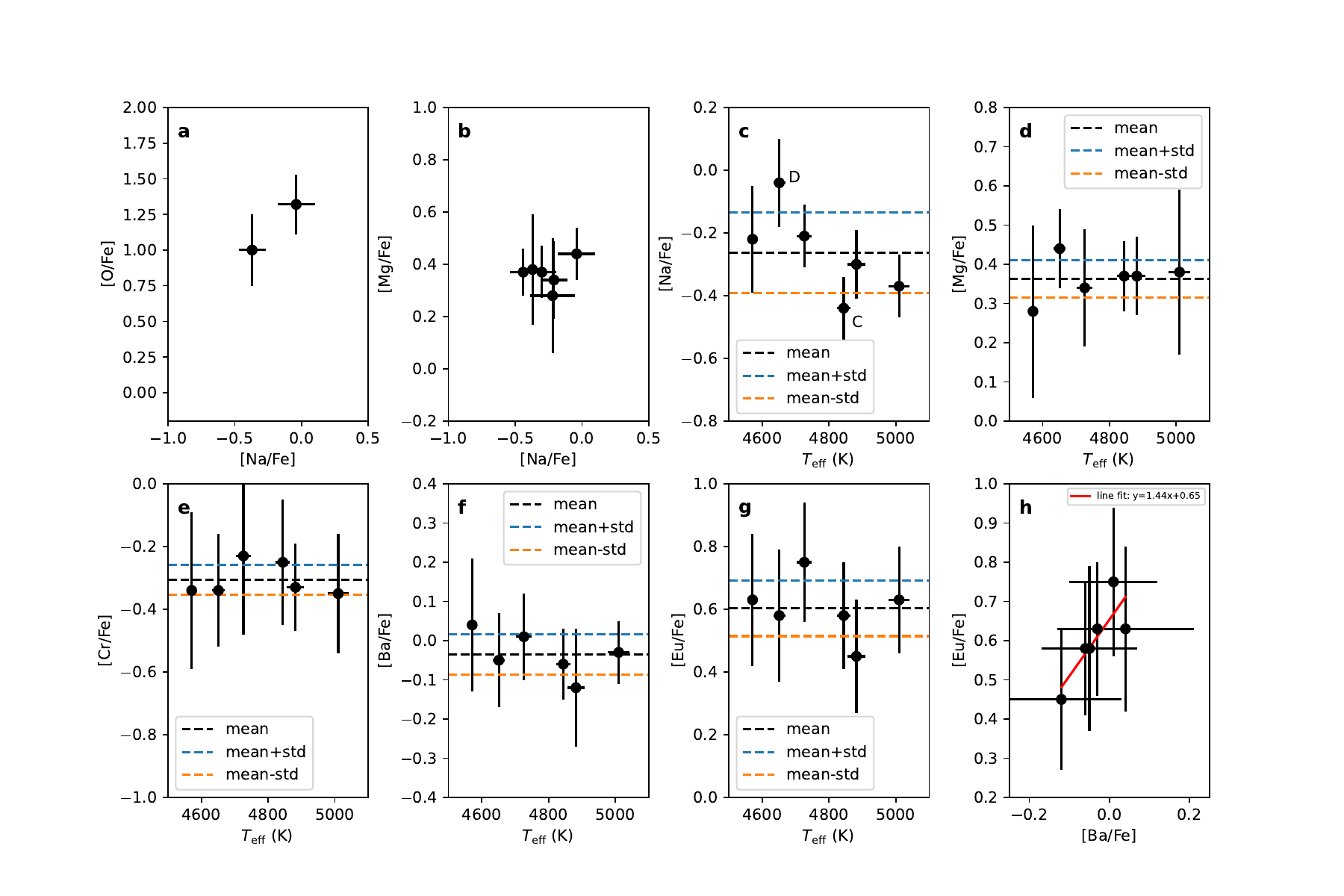}
	\caption{Panel a: [Na/Fe] vs. [O/Fe]. Panel b: [Na/Fe] vs. [Mg/Fe]. Panel c,d,e,f,g: Relation between $T\rm_{eff}$ and [Na/Fe], [Mg/Fe], [Cr/Fe], [Ba/Fe] and [Eu/Fe]. Panel h: The relation between [Ba/Fe]  and [Eu/Fe]. Red line is the line fit. 
		\label{fig:NaO}}
\end{figure*}

\subsection{Memership identification of seven stars with current data }
While LAMOST spectra classify J0930 as a carbon star (\citealt{2016ApJS..226....1J}), its GD-1 membership remained unconfirmed in previous studies. However, its spatial position, proper motion, and chemical abundances ([Fe/H] and [$\rm \alpha/Fe$) strongly support its association with the stream.
The association of J0932 with GD-1 has been verified by  \citep{2019ApJ...877...13H,2019MNRAS.486.2995M,2022MNRAS.515.5802B}. Membership determinations for J0955 and J1015 are supported by \cite{2022MNRAS.515.5802B} and \cite{2018ApJ...869..122L}.  J1048 shows consistent stream alignment across multiple studies \citep{2018ApJ...869..122L,2019ApJ...877...13H,2019MNRAS.486.2995M,2022MNRAS.515.5802B}. The GD-1 membership of J1213 has been confirmed by \cite{2021ApJ...914..123I}, while J1358 was first identified as a stream member in \cite{2018ApJ...869..122L}. This establishes all program stars as confirmed GD-1 members through astrometric and spectroscopic validation.

\subsection{Equivalent width measurement}
The equivalent widths (EWs) are measured from normalized spectra using a dual approach: weak lines (EW $<$ 120 m{\AA}) were fitted with Gaussian profiles, while strong lines (EW $>$ 120 m{\AA}) were measured through direct integration. For the majority of elements, we selected spectral lines with EWs in the optimal range of 10-120 m{\AA}. However, for certain elements with limited available lines (notably Na), we included stronger lines (EW $>$ 120 m{\AA}) in our analysis to ensure adequate spectral coverage and reliable abundance determinations. {\bf The spectral line data and EWs for these seven stars are provided in Table \ref{tab:EW_data}.}

\section{Abundance determination}
Chemical abundances are derived through a standard equivalent width analysis using the ATLAS NEWODF grid of model atmospheres, assuming no convective overshooting (\citealt{2003IAUS..210P.A20C}). The photospheric solar abundances from \citet{2009ARA&A..47..481A} were adopted as reference values when computing the [X/H] and [X/Fe] ratios.

\subsection{Determination of the atmospheric parameters}
The effective temperatures ($T\rm_{eff}$) are derived using the $(V-Ks)\rm_{0}$ color index through empirical calibrations from \citet{1999A&AS..140..261A}, adopting a fixed metallicity of [Fe/H] = -2.30 based on the LAMOST pipeline average (\citealt{2015RAA....15.1095L}). Photometric data were obtained from multiple sources: $Ks$ magnitudes from 2MASS (\citealt{2006AJ....131.1163S}), $V$ magnitudes transformed from Pan-STARRS1 $g$ and $r$ bands (\citealt{2018BlgAJ..28....3K,chambers2019panstarrs1surveys}), with extinction corrections applied using \cite{1998ApJ...500..525S} maps as recalibrated by \cite{2011ApJ...737..103S} (RV = 3.1). $T\rm_{eff}$ uncertainties were propagated from photometric errors. For comparison, we additionally applied two photometric calibrations from \cite{2021A&A...653A..90M} (MBM) using (G-G$\rm_{RP}$) and (G-Ks) color indices.

Since the program stars exhibit parallax uncertainties from Gaia DR3 well above 20$\%$, parallax determinations of surface gravity (log $\rm g_{plx}$) are unreliable. We therefore derive log $g$ using spectroscopy method (log $\rm g_{spa}$), from Fe I/Fe II ionization equilibrium, with uncertainties estimated by requiring the abundance difference to remain within $\pm$0.1 dex. Microturbulent velocities ($\xi_{t}$) are derived by eliminating the dependence of Fe I abundances on equivalent widths, with uncertainties constrained by maintaining a slope of 0.001 in the linear fit. These parameters are iteratively adjusted until consistent solutions were obtained. 

Although \cite{2022ApJ...931..147L} provided a calibration between $T\rm_{eff}$ and the difference (log $\rm g_{spa}$ - log $\rm g_{plx}$)-enabling estimation of log $\rm g_{plx}$ from log $\rm g_{spa}$ and $T\rm_{eff}$. Adopting this log $\rm g_{plx}$ would introduce   Fe I/Fe II abundance discrepancies exceeding 0.2 dex for our seven program stars The log $g$ difference with above calibration for our seven program stars (compared to differences of $\sim$0.3 dex from the calibration). We thus retain our spectroscopically derived log $\rm g_{spa}$ for all subsequent abundance calculations.

Figure \ref{fig:para_cmp} (left panel) compares our $T\rm_{eff}$ determinations with LAMOST pipeline values and alternative calibrations. The MBM (G-G$\rm_{RP}$) and (G-Ks) calibrations yield systematically higher temperatures by 119 K and 98 K respectively, though with small scatters (30 K and 20 K). Compared to LAMOST, we find a systematic offset of 60 K with a scatter of 92 K. For log $g$, the systematic offset is 0.26 dex with 0.21 dex scatter (right panel of Figure 2). Our derived metallicities show remarkable consistency, with mean [Fe/H] = -2.38 $\pm$ 0.05 dex for all seven stars.

Table \ref{tab:Parameter} presents the complete set of observational and derived parameters, including photometric magnitudes ($g, r, i, Ks$), extinction values $E(B-V)$, atmospheric parameters ($T\rm_{eff}$, log $g$, [Fe/H], $\xi_{t}$), radial velocities, and corresponding values from LAMOST and Gaia RVS for comparison.  

\subsection{Elemental Abundance}
The atomic lines are mainly derived from \cite{2013AJ....145...13A}  and \cite{2015ApJ...798..110L}, which covered the wavelength range of 4070$\thicksim$ 6770 {\AA}. The element abundances are calculated with the ABONTEST8 program (Magain et al. private communication), based on the homogeneous, plane-parallel and local thermodynamic equilibrium models of \cite{2003IAUS..210P.A20C}. The program aligns the observed EWs with theoretical predictions derived from atmospheric models, incorporating natural broadening, van der Waals damping broadening, and thermal broadening in the computational process.

Regarding the [X/Fe] ratios of  elements ( Na, O, Mg, Si, Ca, Sc, Ti, Cr, Ni, Sr, Y), the elemental abundances are derived from the measured EWs in conjunction with the stellar parameters listed in Table \ref{tab:Parameter}, while severely blended spectral lines are excluded from the analysis due to their significant uncertainties in abundance determination. For the elements Mn, Ba and Eu, we determine the abundances using the spectral synthesis method and taking into account hyperfine structure (HFS) effect. We accounted for the HFS of  Mn \citep{2011ApJS..194...35D}, Ba \citep{1998AJ....115.1640M} and Eu \citep{2001ApJ...563.1075L} in this work. We utilize the IDL code Spectrum Investigation Utility (SIU) developed by \cite{1999PhDT.......216R} to fit the observed spectra. 

The abundances of these $\alpha$ elements (O, Mg, Si, Ca, and Ti) have been determined for the program stars utilizing:  1 O I line (6300.304 {\AA}), 9-14 Ca I lines, 3-5 Mg I lines, 1 Si I line (4102.936 {\AA}),  14-26 Ti II lines  with well constrained continuum levels placements.

With regard to the light odd Z elements, we measure the abundance of Na from the two NaD lines because other Na lines are too weak.

The abundances determinations of iron-peak elements (Sc, Cr, Mn and Ni) are performed using 4-9 Sc II lines, 3-9 Cr I lines, 2 Mn I lines and 1-3 Ni I lines. For neutron capture elements, the abundances are derived  through 3 Ba II lines for barium; 1 Y II (4398.013 {\AA}) line for yttrium; 1 Eu II line (4129.725 {\AA}) for europium; 1 Sr II line (4215.524 {\AA}) for strontium.

The final stellar abundances in the [X/Fe] are presented in {\bf Table \ref{tab:XFE_uncertainty}}. Notably, our analysis represents the first comprehensive abundance determination for all seven program stars, as no previous measurements are documented in the literature for these targets.

\subsection{Uncertainties}
The determination of stellar abundances involves two primary sources of uncertainty. The first arises from measurement errors in EWs, which are systematically propagated through our abundance calculations. The second originates from uncertainties in fundamental stellar parameters: $T\rm_{eff}$ , log $g$, [Fe/H], and $\xi_{t}$. To quantify these effects, we conducted a parameter sensitivity analysis by individually varying each quantity while maintaining others at their nominal values. 

The resulting abundance variations are presented in Table \ref{tab:XFE_uncertainty}, which detail the impacts of specific perturbations: 2 m{\AA} in EW measurements, 50 K in $T\rm_{eff}$, 0.15 dex in log $g$, 0.1 dex in [Fe/H], and 0.1 km $\rm s^{-1}$ in $\xi_{t}$. Abundance uncertainties account for line-to-line dispersion when multiple transitions are available. For single-line measurements, we assume this uncertainties of 0.15 dex (Sr, Y, Eu) or 0.20 dex (Si). The total uncertainty for each abundance is subsequently calculated as the root sum square of these individual contribution. Our analysis demonstrates that for the majority of chemical elements examined, the combined uncertainties remain below 0.25 dex, confirming the robustness of our abundance determinations against parameter variations.

\section{Result}
The metallicity [Fe/H] of these seven stars are very close. The average value is -2.38 dex with 0.05 dex dispersion, supporting that the GC origin of the GD-1 stream.  The [X/Fe] for 14 elements abundance are showed in column 8 of Table \ref{tab:XFE_uncertainty}. The panels of Figure \ref{fig:abu-dis} demonstrate the abundance pattern of seven program stars for O, Na, Mg, Si, Ca, Sc, Ti, Cr, Mn, Ni, Sr, Y, Ba and Eu. X axis is [Fe/H] while y-axis is [X/Fe]. The gray points are halo metal poor stars from \cite{2022ApJ...931..147L}. The color symbols are member stars of dwarf galaxies Draco, Formax, Ursa Minor and Sextan, which are from SAGA database (\citealt{2008PASJ...60.1159S}). This comparative analysis reveals distinct nucleosynthetic signatures among these populations. The squares in all panels denote the carbon-enhanced star, which exhibits systematically enriched abundances due to binary mass transfer from an evolved companion. Consequently, our subsequent analysis primarily focuses on the remaining six stars that represent the intrinsic chemical properties of the GD-1 stream.

\subsection{\texorpdfstring{$\alpha$}{alpha} elements}
 The observed abundance patterns of $\alpha$ elements (e.g., O, Mg, Si, Ca, Ti) can shed light on the nucleosynthetic processes and star formation histories of stellar populations. The enrichment of $\alpha$ elements is predominantly driven by core-collapse supernovae (Type II supernova, SNe II) from massive stars on short timescales ($\sim$10–100 Myr).

The abundances of $\alpha$ elements Mg, Ca, and Ti exhibit remarkable consistency across the six stars (panels c, e and g of Figure \ref{fig:abu-dis}). The averages of [Mg/Fe], [Ca/Fe] and [Ti/Fe] are 0.36 dex, 0.30 dex and 0.26 dex, respectively. The abundance dispersion of those three elements are within 0.05 dex.  The Si abundance shows a relatively larger scatter (panel d of Figure \ref{fig:abu-dis}). This discrepancy may be attributed to the fact that the Si abundance was determined using only a single Si I line around 4103 {\AA} detectable in our spectra, which might be blended with the wing of H$\delta$.
Thus, we did not take Si as an important element for the abundance analysis.

 The oxygen abundances of the two stars were derived from the O I line at 6300.304 {\AA}. For the remaining four stars, however, the O I lines were severely blended with telluric features, preventing reliable oxygen abundance determinations. While the two measurable stars exhibit a significant difference in [O/Fe] ($\sim$0.32 dex), their combined uncertainties reach up to 0.45 dex. Therefore, no conclusive evidence of oxygen abundance variations is found in this stellar stream.

The Mg abundance is consistent with that of metal-poor stars in the Galactic halo. The dispersion is systematically lower than those observed in dwarf galaxies such as Draco, Ursa Minor, and Sextans (panel c of Figure \ref{fig:abu-dis}). Similarly, the Ca abundance aligns with that of metal-poor halo stars, yet it is notably higher compared to the values found in dwarf galaxies. In contrast, the Ti abundance is lower than that of metal-poor stars in the Galactic halo.

As shown in panels c, d, e and g of Figure \ref{fig:abu-dis}, Ursa Minor, Draco, and Sextans exhibit larger abundance dispersions than both halo metal-poor stars and our program stars. Overall, our program stars show lower Mg and Si abundances compared to dwarf galaxies but higher Ca and Ti abundances.


These discrepancies highlight how $\alpha$ element abundances trace the interplay between star formation efficiency, chemical enrichment timescales, and galactic environment. Dwarf galaxies and halo stars thus represent distinct pathways in hierarchical galaxy assembly.

\subsection{ Sodium}
 For very metal-poor (VMP) and extreme metal-poor (EMP) stars, the Na
 abundance is usually derived from the NaD resonance lines at
 5890 and 5896 {\AA}, and thus usually suffer rather strong NLTE
 effect. For example, the NLTE corrections for Na abundance
 estimated from the NaD lines can be as large as -0.5 dex for
 EMP stars (\citealt{2011A&A...528A.103L}). NLTE corrections for sodium (Na) are available from \cite{2011A&A...528A.103L} via the INSPECT database\footnote{http://www.inspect-stars.com/}. However, as this study focuses on the distribution of six member stars and utilizes LTE-derived Na abundances for the dwarf galaxies, NLTE corrections have not been applied. 
Panel c of Figure \ref{fig:NaO} shows the [Na/Fe] trend along the $T\rm_{eff}$. It is clear that four stars (star A, E, F and G) have almost the same [Na/Fe], while the other two stars (star C and D) demonstrate the large difference, which means [Na/Fe] variation.  Meanwhile, the sodium abundance is consistent with that of metal-poor stars in the Galactic halo (panel f of Figure \ref{fig:abu-dis}),  as demonstrated by \cite{2022ApJ...931..147L}. 

Panel a of Figure \ref{fig:NaO} displays the relationship between [Na/Fe] and [O/Fe]. However, since oxygen abundances were measured for only two stars, a clear Na-O anti-correlation trend could not be established. Panel b shows [Na/Fe] versue [Mg/Fe] relationship, which similarly reveals no significant Na-Mg anti-correlation. This lack of distinct abundance patterns suggests these stars likely belong to a single stellar generation within the GC progenitor.

\subsection{iron-peak elements}

 Iron-peak elements are synthesized in thermonuclear explosions of supernovae (Type Ia supernovae), as well as in
 incomplete or complete Si burning during explosive burning of
 core-collapse supernovae \citep{1995ApJS..101..181W,2006ApJ...653.1145K}.  The abundance patterns of iron-peak elements are presented in Figure \ref{fig:abu-dis} (panels f, h, i, j). Overall, the abundances of Sc, Cr, Mn and Ni are consistent with those of metal-poor stars in the Galactic halo. In contrast, the Cr abundances exhibit minimal scatter, with the exception of J1358, which shows a [Cr/Fe] value of -0.34 dex with an uncertainty of 0.25 dex. Meanwhile, the abundances of Ni, Mn, and Sc display small scatter across the six stars.

However, the NLTE effects for Cr I and Mn I lines in giants cannot be neglected. For example, as demonstrated by \cite{2010A&A...522A...9B},
the NLTE abundance corrections for Cr I lines in metal-poor stars can reach +0.3$\sim$+0.5 dex, with larger deviation at  lower log $g$ and $T\rm_{eff}$.  We calculated the NLTE corrections for Cr I and Mn I lines using the online tool \footnote{https://nlte.mpia.de/gui-siuAC\_secE.php}. The NLTE corrections for Cr I and Mn I exhibit consistency across these six stars, indicating that the absence of NLTE correction would not significantly impact the abundance analysis results.

\subsection{Neutron-capture elements}
 These elements are primarily produced via two distinct mechanisms: the slow (s-) process and the rapid (r-) process. The s-process occurs in low-neutron-density environments, such as the interiors of asymptotic giant branch (AGB) stars, where neutrons are gradually captured over timescales allowing for intermittent radioactive beta decays. This process yields elements like barium. In contrast, the r-process operates under extreme, high-neutron-density events—typically in supernova explosions or neutron star mergers—where nuclei rapidly capture neutrons before undergoing beta decay, creating heavy, neutron-rich isotopes like europium, gold, and uranium.

In this paper, the abundances of neutron-capture elements (Sr, Y, Ba and Eu) in these six stars were determined using  spectral fitting (Ba and Eu) or equivalent width methods (Y and Sr). As shown in the panels k-n of Figure \ref{fig:abu-dis}, their abundance patterns align with those of metal-poor halo stars and fall within the range observed in dwarf galaxies. Among these elements, Ba  exhibit relatively better uniformity across the six stars (panel f of Figure \ref{fig:NaO} ). However, J1048 shows a lower Y abundance compared to the other stars. For Y and Sr, the abundances were derived using only a single spectral line, which may introduce larger uncertainties. The mean abundance ratios of Sr and Y are significantly lower than halo stars and suggests that GD-1 stream's progenitor was born in a lower mass galaxy than most GCs.

The s-process contribution dominates the Galactic evolution of Ba starting from [Fe/H] $>$ -1.5 \citep{1999ApJ...521..691T}. At lower values of [Fe/H], independent of the characteristics of the chemical evolution model, the contribution from s-process nucleosynthesis rapidly decreases due to the strong metallicity dependence of stellar yields. As a result, low-mass AGB stars cannot account for the observed Ba abundances in the range -3 $<$ [Fe/H] $<$ -1.5. In our study, [Fe/H] values of all six stars are below -2.3 dex, indicating that the s-process has a negligible contribution to their Ba abundances.

Panel g of Figure \ref{fig:NaO} shows [Eu/Fe] vs. $T\rm_{eff}$. Five of the six stars exhibit significant europium enrichment with [Eu/Fe] $>$ 0.5 dex, while the remaining star shows [Eu/Fe] = 0.45 dex, clearly indicating their formation in r-process-enriched gas. Interestingly, while all six stars display [Ba/Fe] ratios near solar, panel h of Figure \ref{fig:NaO} reveals a striking correlation between [Eu/Fe] and [Ba/Fe] (Pearson's r = 0.83, p = 0.04) as shown by the linear fit (red line). This strong correlation demonstrates that the barium in these stars must also enriched  from the same r-process event despite its absolute abundance not being enhanced above solar levels. 

\cite{2004ApJ...601..864T} suggested a different stellar origin for Ba and Sr-Y for low metallicity stars. The weak s-process component, occurring during core-He burning and shell-C burning in massive stars \citep{1991ApJ...371..665R}, is primarily responsible for the production of Y and Sr. Both elements exhibit subsolar abundances in our sample, suggesting they were not enriched by the s-process. As shown in panels k, l of Figure \ref{fig:abu-dis}, the Sr and Y abundances in our six stars are systematically lower than those of typical metal-poor halo stars. This indicates that the s-process contributed less to the chemical enrichment of the GD-1 stream progenitor compared to the Milky Way's halo population.

\section{Discussion}
The GD-1 stellar stream has become particularly renowned in galactic dynamics studies due to its exceptionally thin and coherent orbit spanning over 100 degree on the sky. This remarkable geometrical property makes it an exquisite tracer for precisely constraining the Milky Way's gravitational potential. Recent studies have revealed several intriguing substructures along its trajectory, including distinct gaps, density blobs, and spur-like features, which provide crucial diagnostics for studying the perturbative effects of dark matter subhalos during the stream's dynamical evolution. 
 \cite{2025ApJ...980...71V} confirmed a particularly striking discovery a `cocoon' component enveloping the primary stream. This cocoon exhibits substantial physical width (FWHM $\sim$460 pc) and kinematically hot (velocity dispersion, $\sigma$ $\sim$ 5–8 km $\rm s^{-1}$)
component that surrounds  a narrower (FWHM $\sim$ 55 pc) and colder ($\sigma$ = 3.09 $\pm$ 0.76 km $\rm s^{-1}$) thin stream component (based on a  median per star velocity precision of 2.7 km $\rm s^{-1}$). Moreover, they suggested that GD-1 stream might be an accreted GC.

\subsection{metallicity}
The chemical properties of GD-1 stream have been extensively studied in previous works. Based on SEGUE and LAMOST data, the reported [Fe/H] values range from approximately -1.96 to -2.2.  \cite{2025ApJ...980...71V} derived a significantly lower [Fe/H] value of $\sim$-2.54 for this stream using Dark Energy Spectroscopic Instrument (DESI) data , which is much lower than that of previous research.
Our study yields an average [Fe/H] value of -2.38, providing an intermediate result that bridges the gap between previous measurements and the DESI-based estimate. The [Fe/H] dispersion of our seven stars is very small, which is the first confirmation, based on high-resolution spectroscopy, that the GD-1 stream has a uniform metallicity. Thus, our results provide the evidence that GD-1 stream from a GC. 

\subsection{light element abundances}
A few research  reported the $\alpha$ element abundances of the GD-1 stream. \cite{2022MNRAS.515.5802B} selected a sample of likely Red Giant Branch stars from the GD-1 stream for medium-low-resolutionon spectroscopic follow-up and  found 3 $\sigma$ variation in Mg-abundances among the stars in their sample.  Their results suggest that GD-1 represented another fully disrupted low-mass GC where light-element abundance spreads have been found. The chemical properties of our six GD-1 member stars provide intriguing insights into the stream's progenitor system. Panel d of Figure \ref{fig:NaO} shows that the six member stars in our sample exhibit consistent magnesium abundances within measurement uncertainties. However, given the limited sample size, we caution that these results cannot be considered representative of the entire GD-1 stream population. The current data do not provide sufficient evidence to rule out possible magnesium abundance variations along the full extent of the stream. A more comprehensive spectroscopic survey with larger stellar samples would be required to definitively assess the homogeneity of magnesium abundances in GD-1 stream.

While GCs typically exhibit multiple stellar populations characterized by significant light-element variations (e.g., C-N-O, Na-O, and Mg-Al anti-correlations).
Our current dataset reveals several key features:

1. Na-O anti-correlation: Panel a of Figure 4 shows no discernible Na-O anti-correlation trend, though this may reflect observational limitations (only two stars have measurable O abundances).

2. Na abundance dispersion: The mean [Na/Fe] abundance ratio for the six stars is −0.26 dex, with a dispersion of approximately 0.13 dex. This dispersion is notably larger than typical measurement uncertainties. In particular, the [Na/Fe] difference between stars C and D exceeds the combined uncertainties of these two stars. Consequently, the six stars exhibit only weak evidence of intrinsic [Na/Fe] variations.

3. Multiple population evidence: no clear anti-correlations are found in Na-O and Na-Mg.

\subsection{heavy element abundances}
Recent studies of GCs like M2 (\citealt{2014MNRAS.441.3396Y}) and M15 (\citealt{2006ApJ...641L.117O}) have demonstrated that neutron-capture element abundances can vary between cluster members. While our small sample size prevents definitive conclusions, we found  a  correlation between barium and europium (r = 0.83, p = 0.04) that points to a shared r-process origin. Furthermore, the systematically lower abundances of Sr and Y compared to Milky Way halo stars at similar metallicity suggest the progenitor of this stream formed in a low-mass dwarf galaxy environment.

\subsection{abundance comparison of star in main orbit and blob substructure}
Among the six stellar members in our sample, J1015 is uniquely positioned within the blob substructure while the remaining five stars trace the main orbital path of the GD-1 stream. Notably, J1015 exhibits chemical abundance patterns ([X/Fe] ratios) that are fully consistent with those of the other stream members within measurement uncertainties. This chemical homogeneity provides compelling evidence that the blob structure is indeed an authentic component of the GD-1 stream system, rather than a chance alignment of unrelated field stars.  This finding significantly strengthens the interpretation of blob substructures as physically associated overdensities in stellar streams, potentially formed through dynamic interactions with dark matter subhalos or other gravitational perturbations during the stream's orbital evolution.

\section{Conclusion}
Using the LAMOST and SEGUE spectral catalogs, we identified seven member stars of the GD-1 stream for follow-up observations with Subaru/HDS. Detailed stellar parameter estimation and abundance analysis of 14 elements were performed for seven of these stars. The key findings are summarized as follows:

(1) The seven member stars exhibit strikingly consistent [Fe/H] values, providing robust chemical evidence supporting their common origin from a single GC.

(2) Both stars in the `blob' substructure and those in the main orbital track show homogeneous abundance patterns across all measured elements, unambiguously confirming the blob's physical association with the GD-1 stream system.

(3) Notably, our sample displays no significant Mg abundance variations, in contrast to Mg large dispersion reported by \cite{2022MNRAS.515.5802B}. The absence of characteristic Na-O and Na-Mg anti-correlations suggests the progenitor cluster had a relatively low initial mass. 

(4) All six stars with europium measurements show pronounced Eu enhancements ([Eu/Fe] $\sim$ 0.6), indicating the progenitor system formed from gas that had already been enriched by r-process nucleosynthesis events prior to the cluster's formation epoch.

(5) A correlation exists  Eu and Ba abundance, indicating the Ba elements is also enriched by the same r-process event. 

(6)The systematically lower Y and Sr abundances compared to Milky Way halo stars at similar metallicity suggest distinct light neutron-capture element enrichment histories between GD-1's progenitor and the Galactic halo. 

\begin{acknowledgments}
This study is supported by the National Natural Science Foundation of China under grant nos 12588202 and 12273055,  the National Key R\&D Program of China, grant no. 2023YFE0107800 and 2024YFA1611902, the International Partnership Program of Chinese Academy of Sciences under grant No. 178GJHZ2022040GC, the support from the Strategic Priority Research Program of Chinese Academy of Sciences grant No. XDB1160301 and the support by the National Astronomical Observatories, CAS grant no. E4ZB0301. We acknowledge the support from the 2m Chinese Space Station Telescope project. This work has been supported by a Grant-in-Aid for Scientific Research (KAKENHI) (JP22K03688, JP25K01046, JP25HP8011) from the Japan Society for the Promotion of Science.
This research is based on data collected at Subaru Telescope, which is operated by the National Astronomical Observatory of Japan. We are honored and grateful for the
 opportunity of observing the Universe from Maunakea, which has the cultural, historical, and natural significance in Hawaii. Guoshoujing Telescope (the Large Sky Area Multi-Object Fiber Spectroscopic Telescope, LAMOST) is a National Major
 Scientific Project built by the Chinese Academy of Sciences. Funding for the project has been provided by the National Development and Reform Commission. It is operated and managed by the National Astronomical Observatories, Chinese
 Academy of Sciences. This work presents results from the European Space Agency (ESA) space mission Gaia. Gaia data are being processed by the Gaia Data Processing and Analysis Consortium (DPAC). Funding for the DPAC is provided by national institutions, in particular the institutions participating in the Gaia MultiLateral Agreement (MLA). The Gaia mission website is \url{https://www.cosmos.esa.int/gaia}. The Gaia archive website is \url{https://archives.esac.esa.int/gaia}.

\end{acknowledgments}

%




\bibliography{ms}{}
\bibliographystyle{aasjournal}

\appendix
\section{Spectral Line Data and Equivalent Width}
The spectral line data and the EWs of seven stars are listed in Table \ref{tab:EW_data}, which is
 publicly available in its entirety in machine-readable form.
\begin{deluxetable}{ccccccccccc}[ht!]
\tablenum{A1}
\tablecolumns{11}
\label{tab:EW_data}
\tabletypesize{\scriptsize}
\tablecaption{Spectral Line Data and Equivalent Width}
\tablehead{
 \colhead{Wavelength} & \colhead{Species} & \colhead{L.E.P.} & \colhead{log $gf$} & \colhead{J0930} &  \colhead{J0932} & \colhead{J0955} & \colhead{J1015} & \colhead{J1048} & \colhead{J1213} & \colhead{J1358} \\
 \colhead{(\AA)} & \colhead{} & \colhead{(eV)} & \colhead{} & \colhead{(m\AA)} & \colhead{(m\AA)} & \colhead{(m\AA)} & \colhead{(m\AA)} & \colhead{(m\AA)} & \colhead{(m\AA)} & \colhead{(m\AA)}
}
\startdata
4025.129&Ti II&0.61&-2.110&...& 63.9&...&...&...&...&...\\
4028.338&Ti II&1.89&-0.920&...&...& 70.8& 71.0&...&...&...\\
4030.750&Mn I &0.00&-0.497&111.7&...&...&...&...&125.7&...\\
4033.060&Mn I &0.00&-0.647&113.2&104.1&...&...&...&...&...\\
4034.480&Mn I &0.00&-0.843&...&...&...&...&...&110.0&...\\
4041.350&Mn I &2.11& 0.281&...&...& 54.7& 73.0&...& 39.7&...\\
4053.821&Ti II&1.89&-1.070& 24.3& 47.5& 55.6& 59.4&...&...&...\\
4055.540&Mn I &2.14&-0.077&...&...& 31.8& 37.7&...&...&...\\
...&...&...&...&...&...&...&...&...&...&...\\
\enddata
\tablecomments{This table is available in its entirety in machine-readable form.}
\end{deluxetable}

\section{The [X/Fe] and uncertainties of the abundance measurement}
Table \ref{tab:XFE_uncertainty} list the abundance uncertainties and the final [X/Fe] of 14 elements for seven member stars.

\begin{deluxetable}{ccccccccc}[ht!]
\tablenum{B1}
\tabletypesize{\scriptsize}
\setlength{\tabcolsep}{0.02in}
\tablewidth{500pt}
\tablecolumns{9}
\label{tab:XFE_uncertainty}
\tablecaption{ The [X/Fe] and uncertainties of the abundance measurement }
\tablehead{
\colhead{ID}&\colhead{$\Delta$[X/H]} & \colhead{$\frac{\sigma EW}{\sqrt{N}}$}   & \colhead{T$\rm_{eff}$}   & \colhead{log $g$ } &\colhead{[Fe/H]}& \colhead{$\xi_{t}$} &\colhead{[X/Fe]}& \colhead{$\sigma$ Total} \\
  \colhead{}&\colhead{}& \colhead{}  & \colhead{+50K}   & \colhead{+0.15}   & \colhead{+0.1} & \colhead{+0.1} &  \colhead{} & \colhead{}  \\ 
  \colhead{} &\colhead{}& \colhead{}  & \colhead{(K)}   & \colhead{(dex)}   & \colhead{(dex)} & \colhead{(km $\rm s^{-1}$)} &  \colhead{} & \colhead{}  }
\startdata
J0932+2841& BA2  &  0.04  &  0.04  & -0.03  &  0.01  & -0.03  &  -0.03 & 0.08\\
J0932+2841& CA1  &  0.04  &  0.04  &  0.01  &  0.00  & -0.02  &  0.26  & 0.10 \\
J0932+2841& CR1  &  0.05  &  0.07  &  0.01  &  0.00  & -0.04  &  -0.35 & 0.19 \\
J0932+2841& EU2  &  0.04  &  0.04  & -0.04  &  0.00  & -0.02  &  0.63 & 0.17 \\
J0932+2841& FE1  &  0.04  &  0.06  &  0.00  &  0.00  & -0.02  &  --   &  0.15 \\
J0932+2841& FE2  &  0.04  &  0.01  & -0.04  &  0.01  & -0.02  &  --   & 0.12     \\
J0932+2841& MG1  &  0.03  &  0.05  &  0.01  &  0.00  & -0.02  &  0.38 & 0.21 \\
J0932+2841& MN1  &  0.05  &  0.06  &  0.00  & -0.01  & -0.03  &  -0.54 & 0.09    \\
J0932+2841& NA1  &  0.03  &  0.07  &  0.03  &  0.00  & -0.04  & -0.37 &  0.10 \\
J0932+2841& NI1  &  0.04  &  0.05  & -0.01  & -0.01  & -0.02  &  -0.03 & 0.12    \\
J0932+2841& O 1  &  0.14  &  0.03  & -0.04  &  0.00  &  0.00  &  1.0 & 0.25      \\
J0932+2841& SC2  &  0.04  &  0.03  & -0.03  &  0.01  & -0.03  &  0.02 & 0.11 \\
J0932+2841& SI1  &  0.04  &  0.06  &  0.00  &  0.00  & -0.02  &  0.3 & 0.21 \\
J0932+2841& SR2  &  0.04  &  0.06  & -0.01  &  0.01  & -0.06  & -0.19 & 0.17 \\
J0932+2841& TI2  &  0.04  &  0.03  & -0.03  &  0.01  & -0.03  &  0.27 & 0.18 \\
J0932+2841& Y 2  &  0.05  &  0.04  & -0.03  &  0.01  & -0.01  &  -0.11 & 0.17 \\
J0930+2902& BA2  &  0.02  & -0.06  &  0.00  & -0.01  & -0.06  &  1.86 & 0.11 \\
J0930+2902& CA1  &  0.04  & -0.04  & -0.01  &  0.00  & -0.02  &  0.38 & 0.22 \\
J0930+2902& CR1  &  0.05  & -0.06  &  0.00  &  0.00  & -0.03  &  0.02 & 0.22 \\
J0930+2902& FE1  &  0.04  & -0.05  &  0.00  &  0.01  & -0.02  &  --   & 0.12 \\
J0930+2902& FE2  &  0.04  & -0.01  &  0.03  &  0.00  & -0.02  &  --   & 0.11 \\
J0930+2902& MG1  &  0.03  & -0.05  & -0.01  &  0.00  & -0.02  &  0.57 & 0.14 \\
J0930+2902& MN1  &  0.05  & -0.07  & -0.01  &  0.00  & -0.07  &  0.09 & 0.14 \\
J0930+2902& NA1  &  0.02  & -0.04  & -0.02  &  0.01  & -0.01  &  1.61 & 0.11 \\
J0930+2902& NI1  &  0.03  & -0.06  &  0.00  &  0.00  & -0.02  &  -0.09 &0.21  \\
J0930+2902& SC2  &  0.04  & -0.03  &  0.02  &  0.00  & -0.03  &  0.44 & 0.19 \\
J0930+2902& TI2  &  0.04  & -0.03  &  0.03  &  0.00  & -0.03  &  0.41 & 0.21 \\
J0930+2902& Y 2  &  0.03  & -0.05  &  0.01  &  0.00  & -0.07  &  2.40  & 0.13 \\
...&...&...&...&...&...&...&...&...\\
\enddata
\tablecomments{This table is available in its entirety in machine-readable form.}
\end{deluxetable}



\end{document}